\documentclass[final,1p,times,authoryear]{elsarticle}

\makeatletter
\def\ps@pprintTitle{%
 \let\@oddhead\@empty
 \let\@evenhead\@empty
 \def\@oddfoot{\hfill\today}%
 \let\@evenfoot\@oddfoot}
\makeatother

\usepackage{hyperref}
\usepackage{enumitem}
\usepackage{amsmath,amsthm,bm}
\usepackage{cleveref}
\crefname{proposition}{Proposition}{Propositions}

\usepackage{algpseudocode,algorithm}
\algnewcommand\algorithmicinput{\textbf{Input:}}
\algnewcommand\Input{\item[\algorithmicinput]}
\algnewcommand\algorithmicoutput{\textbf{Output:}}
\algnewcommand\Output{\item[\algorithmicoutput]}

\newcounter{savealgorithm}

\usepackage{caption}
\usepackage{mathtools}
\usepackage{comment}
\usepackage{subcaption}
\usepackage{multirow}

\newtheorem{theorem}{Theorem}

\newtheorem{proposition}[theorem]{Proposition}

\theoremstyle{definition}
\newtheorem{definition}[theorem]{Definition}

\newtheorem{remark}{Remark}

\def\cA{{\mathcal A}}
\def\cB{{\mathcal B}}
\def\cC{{\mathcal C}}

\def\cE{{\mathcal E}}

\def\cL{{\mathcal L}}

\def\cP{{\mathcal P}}

\def\cU{{\mathcal U}}
\def\cV{{\mathcal V}}
\def\cW{{\mathcal W}}

\def\JKbasis{{Jordan-Krylov basis}}
\def\JKelim{{Jordan-Krylov elimination}}
\def\KrylovGS{{Krylov generating set}}

\def\ExtKrylovGS{{Extended Krylov generating set}}
\def\extKrylovGS{{extended Krylov generating set}}
\def\KrylovGSProc{{KrylovGS}}
\def\ExtKrylovGSProc{{ExtendedKrylovGS}}
\def\JordanBlockStructureMainProc{{JordanBlocksMain}}
\def\JordanBlockStructureLoopProc{{JordanBlocksLoop}}
\def\JKelimProc{{JordanKrylovElim}}
\def\JordanBlockStructureProc{{JordanBlocks}}
\newcommand{\Z}{\mathbb{Z}}
\newcommand{\K}{K}

\newcommand{\Q}{\mathbb{Q}}
\newcommand{\C}{\mathbb{C}}

\newcommand{\Span}[0]{\mathrm{span}}

\newcommand{\rank}{\mathrm{rank}}

\newcommand{\lbar}{{\bar{\ell}}}

\newcommand{\lhat}{\hat{\ell}}
\newcommand{\add}{\stackrel{\mathrm{\circ}}{\Longleftarrow}}
\newcommand{\getss}[1]{\stackrel{#1}{\Longleftarrow}}

\def\b{{\bm{b}}}
\def\e{{\bm{e}}}
\def\u{{\bm{u}}}
\def\v{{\bm{v}}}
\def\w{{\bm{w}}}
\def\p{{\bm{p}}}

\def\r{{\bm{r}}}

\def\0{{\bm{0}}}
\def\={{\leftarrow}}

\begin{document}

\begin{frontmatter} % elsarticle
  % \title[Calculating the Minimal Annihilating Polynomials of
  % Matrices]
  \title{An Exact Algorithm for Computing the Structure \\of Jordan Blocks%
  % } % Maple Transactions
  \tnoteref{t1}} % elsarticle

  \author[niigata]{Shinichi Tajima}
  \ead{tajima@emeritus.niigata-u.ac.jp} % elsarticle 
  % \author{Shinichi Tajima}
  % \email{tajima@emeritus.niigata-u.ac.jp}
  % \affiliation{%
  % \institution{Niigata University}
  % \city{Niigata}
  % \state{Niigata}
  % \country{Japan}
  % \postcode{950-2181}
% }
  \author[kanazawa]{Katsuyoshi Ohara}
  \ead[url]{http://air.s.kanazawa-u.ac.jp/~ohara/} % elsarticle
  % \author{Katsuyoshi Ohara}
  % \email{ohara@se.kanazawa-u.ac.jp}  
  % \affiliation{%
  % \institution{Kanazawa University}
  % \city{Kanazawa}
  % \state{Ishikawa}
  % \country{Japan}
  % \postcode{920-1192}
% }

  \author[tsukuba]{Akira Terui}\corref{corauthor}%}
  \cortext[corauthor]{Corresponding author} % elsarticle
  \ead[url]{https://researchmap.jp/aterui} % elsarticle
  % \author{Akira Terui}
  % \authornote{Corresponding author}
  % \email{terui@math.tsukuba.ac.jp}
  
  % \affiliation{%
  % \institution{University of Tsukuba}
  % \city{Tsukuba}
  % \state{Ibaraki}
  % \country{Japan}
  % \postcode{305-8571}
% }

  \address[niigata]{Graduate School of Science and Technology, Niigata
    University, Niigata 950-2181, Japan}
  \address[kanazawa]{Faculty of Mathematics and Physics, Kanazawa
    University, Kanazawa 920-1192, Japan}
  \address[tsukuba]{Faculty of Pure and Applied Sciences, University
    of Tsukuba, Tsukuba 305-8571, Japan}

  % elsarticle
  \tnotetext[t1]{This work has been partly supported by JSPS KAKENHI
  Grant Numbers JP18K03320, JP21K03291, JP20K11845, and by the
  Research Institute for Mathematical Sciences, a Joint
  Usage/Research Center located in Kyoto University.}

  % reading abstract
  %\input{eigenvector-2-abstract}
  \begin{abstract}
    %  \begin{linenumbers}
      An efficient method is proposed for computing the structure of Jordan blocks of a matrix of integers or rational numbers by exact computation.
      We have given a method for computing Jordan chains of a matrix with exact computation.
      However, for deriving just the structure of Jordan chains, the algorithm can be reduced 
      to increase its efficiency.
      We propose a modification of the algorithm for that purpose.
      Results of numerical experiments are given.
    %  \end{linenumbers}
  \end{abstract}

  % \begin{CCSXML}
  %   <ccs2012>
  %     <concept>
  %         <concept_id>10010147.10010148.10010149.10010158</concept_id>
  %         <concept_desc>Computing methodologies~Linear algebra algorithms</concept_desc>
  %         <concept_significance>500</concept_significance>
  %     </concept>
  %   </ccs2012>
  % \end{CCSXML}  

  % \ccsdesc[500]{Computing methodologies~Linear algebra algorithms}

  % \maketitle % Maple Transactions

  % \keywords{Minimal annihilating polynomial, Generalized eigenvectors, Jordan chains, Krylov vector space} % Maple Transactions
\end{frontmatter} % elsarticle

\section{Introduction}
\label{sec:intro}

In solving various problems in mathematical sciences, the problem is often reduced to the one of linear algebra. 
Especially in the reduced problem, determining the ``structure'' of the linear transformation on finite dimensional vector spaces may lead to the discovery of the clues in solving the problem, such as the classification of the problem
(\cite{che-del2000,hig2008,wei2022}).
Here, the ``structure'' means the number of Jordan blocks of each size 
associated to an eigenvalue,
with the intention of exploring the qualitative characteristics of the problem.
% In this paper, we propose an efficient method for computing such a structure of the matrix.

We have already proposed an efficient algorithm for computing all the Jordan chains of a square matrix associated to a specific eigenvalue, which also gives its structure (\cite{taj-oha-ter2022}).
While our previous algorithm so far gives comprehensive information on the structure of the matrix, 
% redundancies exist just for calculating the structure of the matrix.
% In particular, 
there exist cases where the structure can be obtained without computing the entire {\JKbasis}.
In this paper, we propose an efficient method for computing just the structure of the matrix by reducing the computation.

Our method has the following features.
First, the entire computation is executed in the field of rational numbers for
efficient computation, while in a simple method, one often 
executes the arithmetic in the algebraic extension.
% of the computable field. 
Second, we do not solve a system of linear equations,
% for computing the Jordan chain, 
while in a simple method, one often 
solves a system of linear equations over the algebraic extension.
Third, we calculate the Jordan chain from the highest rank to the lowest rank,
while in a simple method, 
one usually computes Jordan chains 
by computing generalized eigenvectors from those of lower ranks to 
those of higher ranks.
Instead, we first calculate a ``seed'' vector over $\Q$ 
which corresponds to a generalized eigenvector of the highest rank;
here, the seed vector is called a \emph{\JKbasis}.
We have shown that the entire structure of the matrix is derived from the {\JKbasis}.

The contents of the paper are as follows.
In \Cref{sec:jordan-krylov-basis}, we introduce the notion of the {\JKbasis}
and review the {\JKelim} for computing the {\JKbasis}.
In \Cref{sec:algorithm}, 
based on the computation of generators, which is a base of the {\JKbasis},
and the {\JKelim}, 
we propose an algorithm for computing the structure of the matrix.
In \Cref{sec:complexity-analysis}, we analyze the arithmetic complexity of the proposed algorithm.
In \Cref{sec:experiments}, we give the results of numerical experiments.
In \Cref{sec:concluding-remarks}, we give conclusion of the paper with remarks.

\section{The generalized eigenspace of a matrix and a Jordan-Krylov basis of $\ker f(A)^{\lbar}$}
\label{sec:jordan-krylov-basis}

Let $\C$ be the field of complex numbers and $K\subset \C$ be its computational subfield.
Let $A$ be a square matrix over $\K$ of size $n$, $\pi_A(\lambda)$ the minimal polynomial of $A$,
and $f(\lambda)\in\K[\lambda]$ a monic irreducible factor of $\pi_A(\lambda)$.
Let $\lbar$ be the multiplicity of $f(\lambda)$ in $\pi_A(\lambda)$.
For $1\le\ell\le\lbar$, let
\[
    \ker f(A)^{\ell} = \{\u \in \K^n \mid f(A)^\ell \u = \0\},
\]
with $\ker f(A)^0 = \K^n$.
Then, there exists an ascending chain of subspaces
\[
    \{\0\}\subset\ker f(A)\subset\ker f(A)^2\subset\cdots\subset\ker f(A)^{\lbar}.
\]
Let $\alpha$ be a root of $f(\lambda)$ in $\C$.
The structure of the generalized eigenspace of $A$ associated to $\alpha$ 
corresponds to the structure of $\ker f(A)^{\lbar}$.
For describing the structure of $\ker f(A)^{\lbar}$, 
we introduce the notion of a {\JKbasis} of $\ker f(A)^{\lbar}$.
For $1\le\ell\le\lbar$, if $\u\in\ker f(A)^{\ell}\setminus\ker f(A)^{\ell-1}$, 
then the rank of $\u$ is defined as $\ell$ and denoted by $\rank(\u)=\ell$.

For $\u\in\ker f(A)^{\lbar}$, the vector space
\[
    L_A(\u) = \Span_{\K}\, \{ A^k \u \mid k=0,1,2,\ldots \}  
\]
is called the Krylov subspace.
Since $A$ and $f(A)$ commute, if $\u\in\ker f(A)^{\lbar}$, then
$L_A(\u) \subset \ker f(A)^{\lbar}$.

\begin{definition}[{\KrylovGS, \JKbasis} (\cite{taj-oha-ter2022})]
    Let $W$ be a subspace of $\K^n$. 
    A subset $\cW\subset W$ is called \emph{a {\KrylovGS} of $W$} if it satisfies
    $W=\sum_{\w\in\cW} L_A(\w)$.
    Furthermore, if $\cW$ is finite and satisfies $W=\bigoplus_{\w\in\cW} L_A(\w)$,
    then $\cW$ is called \emph{a {\JKbasis} of $W$}.
\end{definition}

\begin{theorem}[{\cite[Theorem 11]{taj-oha-ter2022}}]
    $\ker f(A)^{\lbar}$ has a {\JKbasis}.    
\end{theorem}

We see the relationship between the {\JKbasis} of $\ker f(A)^{\lbar}$ and the Jordan chains of $A$, 
as follows.
Let $d=\deg f$ and $\alpha_1,\alpha_2,\dots,\alpha_d$ be the roots of $f(\lambda)$ in $\C$.
Let $\psi_f(\mu,\lambda)$ be a symmetric polynomial defined by
\[
    \psi_f(\mu,\lambda) = \frac{f(\mu)-f(\lambda)}{\mu-\lambda}\in\K[\mu,\lambda].
\]
Furthermore, for 
$1\le k\le\lbar$, let $\psi_f^{(k)}(\mu,\lambda)=(\psi_f(\mu,\lambda))^k\mod f(\lambda)$.
Here, ``$\text{mod}\ f(\lambda)$'' means that the coefficients in $(\psi_f(\mu,\lambda))^k$ are
regarded as polynomials in $\K[\lambda]$ and replaced with the remainder of the division
by $f(\lambda)$.

For $\u\in\ker f(A)^{\lbar}$ of rank $\ell$ and $1\le k\le\ell$,
let $\p^{(k)}(\lambda,\u) = \psi_f^{(k)}(A,\lambda E)f(A)^{\ell-k}\u$.
Then, we have the following theorem.

\begin{theorem}
    \label{thm:jordan-chain}
    For $i=1,2,\ldots,d$,
    \begin{equation}\label{eq:J-chain}
        \{ \p^{(\ell)}(\alpha_i,\u), \p^{(\ell-1)}(\alpha_i,\u), \ldots, \p^{(1)}(\alpha_i,\u) \}
    \end{equation}
    gives a Jordan chain of $A$ of length $\ell$ associated to $\alpha_i$ .
\end{theorem}

\Cref{thm:jordan-chain} shows that a vector $\u\in\ker f(A)^{\lbar}$ of rank $\ell$ gives
a representation of a Jordan chain of length $\ell$.
By using a vector in a {\JKbasis} of $\ker f(A)^{\lbar}$ as $\u$, 
the generalized eigenspace of $A$ can be constructed from the {\JKbasis}.
For the Jordan chain in \cref{eq:J-chain}, let $P_A(\alpha_i,\u)$ be the subspace spanned by it as 
\[
    P_A(\alpha_i,\u)=\Span_{\C}\{ \p^{(\ell)}(\alpha_i,\u), \p^{(\ell-1)}(\alpha_i,\u), \ldots, \p^{(1)}(\alpha_i,\u) \}.
\]
For $\cA\subset\ker f(A)^{\lbar}$, 
let $\cA^{(\ell)}=\{\u\in\cA\mid\rank_f\u=\ell\}$
(note that $\cA=\cA^{(1)}\cup\cdots\cup\cA^{(\lbar)}$).
Then, we have the following theorem.

\begin{theorem}[{\cite[Theorem 13]{taj-oha-ter2022}}]
    \label{thm:jkbasis-generalized-eigenspace}
    Let $\cB=\cB^{(\ell)}\cup\cB^{(\ell-1)}\cup\cdots\cup\cB^{(1)}$ be a {\JKbasis} of 
    $\ker f(A)^{\lbar}$.
    Then, for $i=1,2,\ldots,d$, we have the following.
    \begin{enumerate}
        \item A direct sum $\displaystyle \bigoplus_{\b\in \cB^{(\ell)}}P_A(\alpha_i,\b)$ 
        is spanned by Jordan chains of length $\ell$.
        \item A direct sum $\displaystyle \bigoplus_{\b\in \cB}P_A(\alpha_i,\b)$
        gives a generalized eigenspace of $A$ associated to the eigenvalue $\alpha_i$.
        \item $\displaystyle \bigoplus_{\b\in \cB}
        (P_A(\alpha_1,\b)\oplus P_A(\alpha_2,\b)\oplus\cdots\oplus P_A(\alpha_d,\b))
        \simeq \C \otimes_{\K} \ker f(A)^{\bar{\ell}}$.
    \end{enumerate}
\end{theorem}

\Cref{thm:jkbasis-generalized-eigenspace} shows that the structure of the 
Jordan blocks of $A$ associated to $\alpha_i$ which is a root of $f(\lambda)$
is derived from the ranks of the elements and the number of elements of each rank in
the {\JKbasis} of $\ker f(A)^{\lbar}$.
Especially, the elements of rank $\ell$ in the {\JKbasis} correspond to the Jordan chains of length $\ell$.
Thus, we see that the structure of the Jordan blocks of $A$ is derived from the number of elements 
of each rank in the {\JKbasis} of $\ker f(A)^{\lbar}$.

\section{An algorithm for computing the structure of Jordan blocks}
\label{sec:algorithm}

For computing the generalized engenspace of $A$, 
a {\JKbasis} of $\ker f(A)^{\lbar}$ is essentially needed and
it is obtained by an elimination called the {\JKelim} (\cite{taj-oha-ter2022}).
However, the {\JKelim} involves redundant calculation for obtaining the
structure of the Jordan blocks.
Thus, we investigate how to reduce a redundant part of the {\JKelim} 
for computing the structure of the Jordan blocks.
Note that, although the {\JKbasis} is not necessarily unique, 
the number of elements of each rank in the {\JKbasis} is unique.

\subsection{Calculating {\KrylovGS}}
\label{sec:calc-krylovGS}

{\JKbasis} is calculated from the {\KrylovGS} of $\ker f(A)^{\lbar}$ defined 
as above.
{\KrylovGS} are calculated using the minimal annihilating polynomials 
of the vectors in a basis of $\K^n$.
Assume that the characteristic polynomial of $A$ is given as 
\begin{equation}
  \label{eq:chara-pol}
  \chi_A(\lambda)=f(\lambda)^{m}g(\lambda),
\end{equation}
where $f(\lambda)$ is a monic irreducible polynomial in $\K[\lambda]$ and 
$f(\lambda)$ and $g(\lambda)$ are relatively prime.

\begin{definition}[The minimal annihilating polynomial]
  For $\u\in\K^n$, 
  let $\pi_{A,\u}(\lambda)$ be the monic generator of a principal ideal 
  $\mathrm{Ann}_{K[\lambda]}(A,\u)=\{g(\lambda)\in K[\lambda] \mid g(A)\u = \0\}$.
  Then, $\pi_{A,\u}(\lambda)$ is called the \emph{minimal annihilating polynomial of $\u$ with respect to $A$}.
\end{definition}

Let $\cE$ be a basis in $\K^n$ and $\cP=\{\pi_{A,\e}(\lambda) \mid \e\in\cE\}$.
For $\e\in\cE$, the minimal annihilating polynomial $\pi_{A,\e}(\lambda)$ is calculated by
\begin{equation}
  \label{eq:min-annihilating-polynomial}
  \pi_{A,\e}(\lambda)=f(\lambda)^{\ell_\e}g_\e(\lambda), \quad \gcd(f,g_\e)=1,\quad
  \ell_\e\ge 0,
\end{equation}
thus we have 
$\lbar=\max\ \{ \ell_\e \mid \e \in \cE\}$
and $g_\e(A)\e\in \ker f(A)^{\lbar}$.
Let 
\begin{equation}
  \label{eq:krylovGS}
  \cE_f=\{\e\in\cE\mid \pi_{A,\e}(\lambda)=f(\lambda)^{\ell_\e}g_\e(\lambda),\ 
  \ell_{\e}>0\},
  \quad
  \cV = \{ g_\e(A)\e \mid \e\in \cE_f\}.
\end{equation}
Then, for $\u\in\cV$, we have $L_A(\u)\subset\ker f(A)^{\lbar}$.

Assume that the rank of $\u\in\ker f(A)^{\lbar}$ is equal to $\ell$, 
and let
\begin{gather*}  
  \cL_{A,d}(\u)=\{\u,A\u,A^2\u,\dots,A^{d-1}\u\},
  \\
  \cL_A(\u)=\cL_{A,d}(\u)\cup\cL_{A,d}(f(A)\u)\cup\cdots\cup\cL_{A,d}(f(A)^{\ell-1}\u).
\end{gather*}
Then, we have $L_A(\u)=\Span_{\K}\cL_A(\u)$.
For $\cU\subset\K^n$, Let $\cL_A(\cU)=\bigcup_{\u\in\cU}\cL_A(\u)$.

\begin{proposition}
  \label{prop:ker-f(A)^l}
  $\ker f(A)^{\lbar}=\Span_{\K}\cL_A(\cV)$.
\end{proposition}

Proposition \ref{prop:ker-f(A)^l} shows that $\cV$ is a {\KrylovGS} set of $\ker f(A)^{\lbar}$.

Now, we investigate the structure of the Jordan blocks of $A$ 
associated to the eigenvalue $\alpha\in\C$ that is a root of 
$f(\lambda)$ in \cref{eq:chara-pol}.
For $\e\in\cE$, assume that the minimal annihilating polynomial 
$\pi_{A,\e}(\lambda)$ is given as
\begin{equation}
  \label{eq:pi_a_ej_f1}
  \pi_{A,\e}(\lambda)=f(\lambda)^{\ell_\e}g_\e(\lambda), \quad \gcd(f,g_\e)=1,\quad
  \ell_\e\ge 0,
\end{equation}
then the minimal annihilating polynomial of $f(A)^{m}\e$ is $g_{\e}(\lambda)$.
Thus, by \cref{eq:pi_a_ej_f1}, we have $g_{\e}(A)\e\in\ker f(A)^{\ell_{\e}}$.
By setting
\begin{equation}
  \label{eq:l1bar}
  \lbar=\max\{\ell_{\e}\mid \e\in\cE\},\quad
  \cV=\{g_{\e}(A)\e\mid \e\in\cE\},  
\end{equation}
$\cV$ is a {\KrylovGS} of $\ker f(A)^{\lbar}$.

\begin{algorithm}[t]
  \caption{Computing a {\KrylovGS} of $\ker f(A)^{\lbar}$}
  \label{alg:KrylovGS}
  \begin{algorithmic}[1]
    \Input{
      A matrix $A\in\K^{n\times n}$,
      the characteristic polynomial 
      $\chi_A(\lambda)=f(\lambda)^{m}g(\lambda)$
      expressed as in \eqref{eq:chara-pol},
      an irreducible factor $f(\lambda)$,
      a basis $\cE=\{\e_1,\dots,\e_n\}$ of $\K^n$ 
    }
    \Output{
      a {\KrylovGS} $\cV$ of $\ker f(A)^{\lbar}$
      % $\{\tilde\cV,\lbar\}$, where
      % $\tilde\cV=\{(\v, \rank_{f_1}\v, f_1(A)^{\rank_{f_1}\v-1}\v)\}$:
      % an \ExtKrylovGS\ $\tilde{\cV}$  of $\ker f_1(A)^{\lbar}$,
      % % An \ExtKrylovGS\ $\tilde{\cV}$  of $\ker f_1(A)^{\lbar}$,
      % $\lbar=\max\{\rank_{f_1}\v\mid\v\in\cV\}$
    }
    \Function{\KrylovGSProc}{$A$, $\chi_A(\lambda)$, $f(\lambda)$, $\cE$}
      \State{$\cV\gets\emptyset$}
      \For{$i=1,\dots,n$}
        \State{$\e'_i\gets f(A)^{m}\e_i$}
        \label{line:alg:KrylovGS:e'_i=0}
        \If{$\e'_i=\0$} 
          \State{$g_i(\lambda)\gets 1$} 
        \Else
          \State $g_i(\lambda)\gets\text{(the minimum annihilating polynomial of $\e'_i$ (\cite{taj-oha-ter2018}))}$
          \label{line:alg:KrylovGS:g_i}
        \EndIf
        \State{$\v_i\gets g_i(A)\e_i$} \label{line:alg:KrylovGS:v_i}
        \label{line:alg:KrylovGS:v_i}
        \State{$\cV\gets\cV\cup\{\v_i\}$ \textbf{if} $\v_i\ne\0$}
      \EndFor 
      \label{line:alg:KrylovGS:endfor}
      \State \Return $\cV$
      % \State $\{\tilde{\cV},\lbar\}\gets\textproc{ConstructExtKrylovGS}(A, f_1(\lambda), \cV)$;
      % \State \Return $\{\tilde{\cV},\lbar\}$
    \EndFunction
  \end{algorithmic}
\end{algorithm}
\begin{algorithm}[H]
  % \ContinuedFloat
  \caption{Computing the {\ExtKrylovGS} of $\ker f(A)^{\lbar}$}
  \label{alg:ExtKrylovGS-matrix}
  \begin{algorithmic}[1]
    \Input{
      A matrix $A\in\K^{n\times n}$,
      % The characteristic polynomial 
      % $\chi_A(\lambda)=f_1(\lambda)^{m_1}\cdots f_k(\lambda)^{m_k}$ 
      % \eqref{eq:chara-pol},
      an irreducible factor $f(\lambda)$,
      % A basis $\cE=\{\e_1,\dots,\e_n\}$ of $\K^n$ 
      a {\KrylovGS} $\cV$ of $\ker f(A)^{\lbar}$
    }
    \Output{
      $\{\tilde\cV,\lbar\}$, where 
      $\tilde\cV=\{(\v, \rank_{f}\v, f(A)^{\rank_{f}\v-1}\v)\}$:
      an \ExtKrylovGS\ $\tilde{\cV}$  of $\ker f(A)^{\lbar}$,
      % $\ker f_1(A)^{\lbar}$の \ExtKrylovGS,
      $\lbar=\max\{\rank_{f}\v\mid\v\in\cV\}$
    }
    \Function{\ExtKrylovGSProc}{$A$, $f(\lambda)$, $\cV$}
      \State{$\tilde{\cV}\gets\emptyset;\quad\lbar\gets 0$}
      % \quad $\cE'\gets\{\e'_i=f_1(A)^{m_1}\e_i\mid \e_i\in\cE, i=1,\dots,n\}$;
      % \State $\cV\gets\textproc{KrylovGS($A$, $\chi_A(\lambda)$, $f_1(\lambda)$, $\cE$)}$;
      % \Comment{\Cref{alg:KrylovGS}}
      \State{(Optional) column reduction: $[\cV]\longrightarrow[\cV']$}
      \State{$[\cV'_0]\gets[\cV']$, \quad $\ell\gets 0$, 
      \quad $N\gets\mbox{(the number of columns in $[\cV']$)}$}
      \While{$[\cV']\ne O$}
        \label{line:alg:ExtKrylovGS-matrix:begin-while}
        \State{$[\cV'']\gets f(A)[\cV']$,\quad $\ell\gets\ell+1$,\quad $\lbar\gets\ell$}
        \For{$j=1,\dots,N$}
          \State $\tilde{\cV}\add([\cV'_0]_j,\ell,[\cV']_j)$ 
          \textbf{if} $[\cV'']_j=\0$ and $[\cV']_j\ne\0$
        \EndFor
        \State $[\cV']\gets[\cV'']$
      \EndWhile
      \label{line:alg:ExtKrylovGS-matrix:end-while}
      \State \Return $\{\tilde{\cV},\lbar\}$
      % \label{line:alg:ExtKrylovGS:return}
    \EndFunction
  \end{algorithmic}
\end{algorithm}

In calculating $\v\in\cV$, we calculate $\rank_f\v$ and 
$\v'= f(A)^{\rank_{f}\v-1}\v$. 
Since these are to be used for calculating the structure of the Jordan blocks of $A$,
we keep these values for future use as 
$\tilde{\cV}=\{(\v,\rank_{f}\v, f(A)^{\rank_{f}\v-1}\v)\mid\v\in\cV\}$,
which is called the \emph{{\ExtKrylovGS}} of $\ker f(A)^{\lbar}$.
Algorithms for computing a {\KrylovGS} and the extended {\KrylovGS}
are given in \Cref{alg:KrylovGS,alg:ExtKrylovGS-matrix}, respectively.
Note that the output of \Cref{alg:KrylovGS} is the input of 
\Cref{alg:ExtKrylovGS-matrix}.
In \Cref{alg:ExtKrylovGS-matrix}, 
adding $(\v,\rank_{f}\v, f(A)^{\rank_{f}\v-1}\v)$ into 
$\tilde{\cV}$ is denoted by 
$\tilde{\cV}\add (\v,\rank_{f}\v, f(A)^{\rank_{f}\v-1}\v)$,
with changing the structure of $\tilde{\cV}$.
Extracting the subset of $\tilde{\cV}$ with the rank $\ell$ is denoted by
$\tilde{\cV}^{(\ell)}=
\{(\v,\rank_{f}\v, f(A)^{\rank_{f}\v-1}\v)\in\tilde{\cV}\mid \rank_{f}\v=\ell\}$.
For a set of finite column vectors $\cW=\{\w_1,\w_2,\dots,\w_k\}$, 
the matrix consisting of all the vectors in $\cW$ is denoted by 
$[\ \cW\ ]=[\w_1,\w_2,\dots,\w_k]$.
For a matrix $M$, the $j$-th column is denoted by $M_j$.

\begin{proposition}
  \Cref{alg:KrylovGS,alg:ExtKrylovGS-matrix} output an {\extKrylovGS} of 
  $\ker f(A)^{\lbar}$.
\end{proposition}
% \begin{proof}
%   We first show that the output of \Cref{alg:KrylovGS} is a 
%   {\KrylovGS} of $\ker f_1(A)^{\lbar_1}$.
%   If $\e_i'=\0$ in line \ref{line:alg:KrylovGS:e'_i=0}, 
%   we see that $\v_i=g_{i}(A)\e_i=\e_i$ in line \ref{line:alg:KrylovGS:v_i};
%   otherwise $\v_i=g_{i}(A)\e_i$.
%   Thus, the output $\cV$ equals $\cV$ in \cref{eq:l1bar}.
% \end{proof}

\subsection{Computing the structure of Jordan blocks}
\label{sec:computing-structure-jordan-blocks}

Let $\cB=\bigcup_{\ell=1}^{\lbar}\cB^{(\ell)}$ be
a {\JKbasis} of $\ker f(A)^{\lbar}$.
The {\JKbasis} is calculated from the {\KrylovGS} $\cV$ as follows.
First, let the first element in $\cB^{(\lbar)}$ be 
$\u\in\cV^{(\lbar)}$ of the highest rank in $\cV$.
Then, the other elements in $\cB^{(\lbar)}$ are calculated by
elimination of the elements in $\cV^{(\lbar)}$.
After that, for $\ell=\lbar-1,\lbar-2,\dots,1$, 
the elements in $\cB^{(\ell)}$ are calculated by elimination
of the elements in $\cV^{(\ell)}$.
The elimination used in this process involves the Krylov vector space
and is called the {\JKelim} (\cite{taj-oha-ter2022}).

For $i=1,2,\ldots,d$, let $c_i=|\cB_i|$,
and call 
$\cC=\{c_1,c_2,\dots,c_d\}$ the \emph{structure of the Jordan blocks}.
In computing $\cC$, we reduce the number of 
{\JKelim} to the number of times possible required to find $c_i$. 
For $m$ in \cref{eq:chara-pol}, we have
\begin{equation}
  \label{eq:m}
  m=\sum_{\ell=1}^{\lbar}\ell\cdot |\cB^{(\ell)}|,
\end{equation}
since $m_1$ is equal to the sum of the lengths of the Jordan chains of $A$
associated to the eigenvalue $\alpha$, whose generators are linearly 
independent over $\C$.
\Cref{eq:m} implies the following proposition.

\begin{proposition}
  \label{prop:rank<k}
  If $m-\sum_{\ell=k+1}^{\lbar}\ell\cdot |\cB^{(\ell)}|<k$,
  then $\cB^{(k)}=\emptyset$.
\end{proposition}
Proposition \ref{prop:rank<k} tells us that, if 
$m - \sum_{\ell=k+1}^{\lbar}\ell\cdot |\cB^{(\ell)}|\le 1$, 
then the structure of the Jordan blocks is determined.
This gives an idea for computing the structure of the Jordan blocks as 
follows.
Let $k\ge 1$. 
After computing {\JKelim} for rank $\lbar_1,\lbar_1-1,\dots,k+1$,
if $m'=m-\sum_{\ell=k+1}^{\lbar_1}\ell\cdot c_{\ell}<k$, 
then perform {\JKelim} for rank $m'$ first. 
If $\cB^{(m')}=\emptyset$, then continue to perform {\JKelim} for rank $k$ 
and below, and calculate 
\begin{equation}
  \label{eq:undecided-multiplicity}
  m-\sum_{\ell=k}^{\lbar}\ell\cdot |\cB^{(\ell)}|.
\end{equation}

An algorithm for computing the structure of the Jordan blocks is 
shown in \Cref{alg:jordan-blocks-structure-main-alt}.
The loop performed in 
\Cref{alg:jordan-blocks-structure-main-alt} is shown in
\Cref{alg:jordan-blocks-structure-loop-alt}, and 
the {\JKelim} performed in \Cref{alg:jordan-blocks-structure-loop-alt}
is shown in \Cref{alg:Jordan-Krylov-elim-ell-alt}.
The overall algorithm for computing the structure of the Jordan blocks of $A$
associated to the root of $f(\lambda)$ is shown in \Cref{alg:jordan-blocks-structure}.
In the algorithms, rewriting the $\ell$-th element in an ordered set 
(such as $\cB$) is denoted by $\getss{\ell}$. 
% For a matrix $M$, the sub-matrix consisting of columns from $s$ to $t$ is 
% denoted by $M_{s..t}$.  

\begin{remark}
  \label{rem:undecided-multiplicity}
  In \Cref{alg:Jordan-Krylov-elim-ell-alt}, the sum of undetermined numbers and sizes 
  of the Jordan blocks as shown in \cref{eq:undecided-multiplicity} is called
  ``the undetermined multiplicity of $f$.''
\end{remark}

\begin{algorithm}[t]
  % \ContinuedFloat
  \caption{Computing the structure of Jordan blocks}
  \label{alg:jordan-blocks-structure-main-alt}
  \begin{algorithmic}[1]
    \Input{A matrix $f(A)\in\K^{n\times n}$,
    % an irreducible factor $f(\lambda)\in\K[\lambda]$,
    an {\extKrylovGS} 
    $\tilde{\cV}$
    % =\bigcup_{\ell=1}^{\lbar}\tilde{\cV}^{(\ell)}$ 
    of $\ker f(A)^{\lbar}$,

    the multiplicity $m$ of $f(\lambda)$ in the characteristic polynomial of $A$}
    \Output{The structure of the Jorndan Blocks of $A$ associated to 
    a root of $f(\lambda)$ $\cC=\{c_1,c_2,\dots,c_{\lbar}\}$}
    \Function{\JordanBlockStructureMainProc}{$f(A)$, $\tilde{\cV}$, $m$}
      \State{\Return $\{m,0,\dots,0\}$ \textbf{if} $\lbar=1$}
      \State{$m\gets m-\lbar$}
      \State{\Return $\{m,0,\dots,0,1\}$ \textbf{if} $m\le 1$}
      \Comment{The structure of the Jordan blocks is determined for $m=0$ or $1$}
      \State{$\cC\gets \{0,\dots,0\}$,\quad 
        $S=\{S_2,S_3,\dots,S_{\lbar}\}\gets\{0,\dots,0\}$,\quad
        $\cB=\{\cB^{(2)},\dots,\cB^{(\lbar)}\}\gets\{\emptyset,\dots,\emptyset\}$
      } 
      \Comment{Initialization of $\cC$, $S$, $\cB$}
      % \label{line:alg:jordan-blocks-structure:lbar=1}
      \State{$\lhat\gets\lbar$} 
      \Comment{The maximum value of the rank of which the {\JKelim} is incompleted}
      \State{$N\gets|\tilde{\cV}^{(\lbar)}|$}
      \State{Choose $(\v,\lbar,\v')\in\tilde{\cV}^{(\lbar)}$, \quad
      $\cB\getss{\lbar}\{\v\}$,\quad
      $\tilde{\cV}^{(\lbar)}\gets\tilde{\cV}^{(\lbar)}\setminus\{(\v,\lbar,\v')\}$,\quad
      $\cC\getss{\lbar}1$
      } 
      % \label{line:alg:jordan-blocks-structure:lbarblock}
      \Comment{A Jordan block of rank $\lbar$ is detected}
      % \If{$m\le 1$} \label{line:jordan-krylov-basis:v}
      %   \State{$\cC[1]\gets m$}
      %   \State{\Return $\cC$} 
      % \EndIf
      \State $S\getss{\lbar}[\cL_{A,d}(\v)]$ 
      % \label{line:alg:jordan-blocks-structure:defineslbar}
      \Comment{Keep $S$ by rank}
      \State $W\gets [\cL_{A,d}(\v')]$
      % \State $W\gets f(A)^{\lbar-1}S_{\lbar}$
      \Comment{$W=f(A)^{\lbar-1}S_{\lbar}$ can be reduced}
      \State{$\ell\gets\min\{\lhat,m\}$} 
      % \label{line:alg:jordan-blocks-structure:minl-1}
      \State {$\cC\gets\textproc{\JordanBlockStructureLoopProc}$ ($f(A)$, $\ell$, $\lhat$, $m$, $S$, $W$, $\tilde{\cV}$, $\cB$, $\cC$)}
      \Comment{\Cref{alg:jordan-blocks-structure-loop-alt}}
      % ランク$1$では$m$がJordan細胞の個数に等しい}
      % \State{$\cC\getss{1}m$}       
      \State{\Return $\cC$}
    \EndFunction
  \end{algorithmic}
\end{algorithm}

\begin{algorithm}
  % \ContinuedFloat
  \caption{Computing the structure of Jordan blocks (loop in $\ell$)}
  \label{alg:jordan-blocks-structure-loop-alt}
  \begin{algorithmic}[1]
    \Input{
      $f(A)\in\K^{n\times n}$: a matrix ,
      $\ell$: current rank,

      $\lhat$: The maximum value of the rank of which the {\JKelim} is incompleted,

      $m$: the undecided multiplicity of $f(\lambda)$ (see \Cref{rem:undecided-multiplicity}),
      
      $S$, $W$: matrices used in \JKelim,

      $\tilde{\cV}$: an {\ExtKrylovGS} of $\ker f(A)^{\lbar}$,

      $\cB$: a temporary {\JKbasis} of $\ker f(A)^{\lbar}$,

      $\cC=\{c_1,c_2,\dots,c_{\lbar}\}$: the structure of the Jordan blocks
    }
    \Output{
      $\cC=\{c_1,c_2,\dots,c_{\lbar}\}$: the structure of the Jordan blocks
    }
    \Function{\JordanBlockStructureLoopProc}{$f(A)$, $\ell$, $\lhat$, $m$, $S$, $W$, $\tilde{\cV}$, $\cB$, $\cC$}
      \While{$\ell>1$} 
        % \label{line:alg:jordan-blocks-structure:while-loop}
        \State{$S\getss{\ell} f(A)^{\lhat-\ell}S_{\lhat}$ \textbf{if} $S_{\ell}=0$ and $\ell<\lhat$}
        % \label{line:alg:jordan-blocks-structure:updatesl}
        \Comment{If $S_{\ell}$ is not calculated yet, calculate it at this step}
        \State{$\{m,S,W,\tilde{\cV},\cB,\cC\}$ $\gets$ \textproc{\JKelimProc}($f(A)$, $\ell$, $m$, $S$, $W$, $\tilde{\cV}$, $\cB$, $\cC$)} 
        % \label{line:alg:jordan-blocks-structure:jkelim-l}
        \Comment{The {\JKelim} of rank $\ell$ (\Cref{alg:Jordan-Krylov-elim-ell-alt})}
        \State{\Return $\cC$ \textbf{if} $m\le 1$}
        \Comment{The structure of the Jordan blocks is determined for $m=0$ or $1$}  
        \State{$S\getss{\ell-1} f(A)S_{\ell}$}
        \State{$\lhat\gets\ell-1$ \textbf{if} $\lhat=\ell$}          
        % \label{line:alg:jordan-blocks-structure:updatelhat-1}
        \If{$\cB_{\ell}=\emptyset$ and $\ell<\lhat$} 
          % \label{line:alg:jordan-blocks-structure:updatejkelim}
          \Comment{Perform the {\JKelim} for $\lhat,\lhat-1,\dots,\ell$}
          \For{$\ell'=\lhat,\lhat-1,\dots,\ell$}
            % \label{line:alg:jordan-blocks-structure:for-loop}
            \State{$\{m,S,W,\tilde{\cV},\cB,\cC\}\gets$ \textproc{\JKelimProc}($f(A)$, $\ell'$, $m$, $S$, $W$, $\tilde{\cV}$, $\cB$, $\cC$)}
            \Comment{The {\JKelim} of rank $\ell'$ (\Cref{alg:Jordan-Krylov-elim-ell-alt})}
            \State{\Return $\cC$ \textbf{if} $m\le 1$}
            \Comment{The structure of the Jordan blocks is determined for $m=0$ or $1$}  
            \State{$S\getss{\ell-1} f(A)S_{\ell'}$}
          \EndFor
          \State{$\lhat\gets\ell-1$}
          % \label{line:alg:jordan-blocks-structure:updatelhat-2}
        \EndIf
        \State{$\ell\gets\min\{m,\ell-1\}$}
        % \label{line:alg:jordan-blocks-structure:minl-2}
        \Comment{The next rank of which to perform {\JKelim}}
      \EndWhile
    \State{$\cC\getss{1}m$}
    \State{\Return $\cC$}
    \EndFunction
  \end{algorithmic}
\end{algorithm}

\begin{algorithm}[t]
  % \ContinuedFloat
\caption{The {\JKelim} of rank $\ell$}
\label{alg:Jordan-Krylov-elim-ell-alt}
\begin{algorithmic}[1]
  \Input{
    A matrix $f(A)\in\K^{n\times n}$, 
    % an irreducible factor $f(\lambda)\in\K[\lambda]$,
    A rank $\ell$ of $\ker f(A)^{\lbar}$},

  an undecided multiplicity of $f(\lambda)$ (see \Cref{rem:undecided-multiplicity})
  $m$, 

  matrices used in {\JKelim} $S$, $W$,     

  an {\ExtKrylovGS} of $\ker f(A)^{\lbar}$ $\tilde{\cV}$,

  a temporary {\JKbasis} of $\ker f(A)^{\lbar}$ $\cB$,
    
  the structure of Jordan blocks $\cC=\{c_1,c_2,\dots,c_{\lbar}\}$
  \Output{
    $\{m,S,W,\tilde{\cV},\cB,\cC\}$, where
    $m$: an undecided multiplicity of $f(\lambda)$ (see \Cref{rem:undecided-multiplicity}), 

    $S$, $W$: matrices used in {\JKelim},

    $\tilde{\cV}$: an extended {\KrylovGS},
    $\cB$: a temporary {\JKbasis}, 

    $\cC=\{c_1,c_2,\dots,c_{\lbar}\}$: the structure of Jordan blocks 
  }

  \Function{\JKelimProc}{$f(A)$, $\ell$, $m$, $S$, $W$, $\tilde{\cV}$, $\cB$, $\cC$}
  \While {$\tilde{\cV}^{(\ell)} \ne \emptyset$}
    \State{$N\gets(\text{the number of columns in }[\tilde{\cV}^{(\ell)}])$ }
    \State Choose $(\v,\ell,\v')\in\tilde{\cV}^{(\ell)}$, \quad
      $\tilde{\cV}^{(\ell)}\gets\tilde{\cV}^{(\ell)}\setminus\{(\v,\ell,\v')\}$
      \Comment{$\v'=f(A)^{\ell-1}\v$}
    % \State $\v'\gets f(A)^{\ell-1}\v$
    \State Simultaneous column reduction of the rightmost column in the augmented matrix 
    
    \qquad
    $[W\mid\v']\longrightarrow[W\mid\r']$,\quad $[S_{\ell}\mid\v]\longrightarrow[S_{\ell}\mid\r]$
    \If{$\r' \ne \0$}
    \Comment{$\r'\not\in\Span_K W$}
      \State $\cB\getss{\ell}\{\r\}$, \quad $m\gets m-\ell$, \quad 
      $\cC\getss{\ell} \cC_{\ell}+1$
      \Comment{The number of Jordan Blocks of rank $\ell$ increases by $1$}
      \If{$m\le 1$} % \label{line:jordan-krylov-basis:ell}
      \Comment{The structure of the Jordan blocks is determined for $m=0$ or $1$}
      %   \State{\Return{$\cB=\cB^{(\lbar)}\cup\cB^{(\lbar-1)}\cup\cdots\cup\cB^{(\ell)}$}}
          \State{$\cC\getss{1} m$}
          \State{\Return $\{m,S,W,\tilde{\cV},\cB,\cC\}$} 
      \EndIf
      \State
          $S\getss{\ell}[S_{\ell}\mid\cL_{A,d}(\r)]$,\quad 
          $W\gets[W\mid\cL_{A,d}(\r')]$
          % \State $S\gets S'$,\quad $W\gets W'$
          \Comment{$W=f(A)^{\ell-1}S$ can be reduced}
          \ElsIf {$\r\ne \0$ and $\ell>1$}
          \State $\ell'\gets\rank_f\r$,\quad 
          $\tilde{\cV}^{(\ell')}\add(\r,\ell',\r')$
    \EndIf        
  \EndWhile
  \State{\Return $\{m,S,W,\tilde{\cV},\cB,\cC\}$}
  \EndFunction
\end{algorithmic}
\end{algorithm}

\begin{algorithm}[t]
  \caption{Computing the structure of Jordan blocks of $A$ associated to
  the root of $f(\lambda)$}
  \label{alg:jordan-blocks-structure}
  \begin{algorithmic}[1]
    \Input{
      a matrix $A\in\K^{n\times n}$,
      the characteristic polynomial
      $\chi_A(\lambda)=f(\lambda)^{m}g(\lambda)$ 
      expressed as in \eqref{eq:chara-pol},
      an irreducible factor $f(\lambda)$,
      a basis $\cE=\{\e_1,\dots,\e_n\}$ of $\K^n$  
    }
    \Output{The structure of Jordan blocks of $A$ 
      associated to the root of $f(\lambda)$: $\cC=\{c_1,c_2,\dots,c_{\lbar}\}$,
     where $\lbar$ is the output of \Cref{alg:ExtKrylovGS-matrix}}
    \Function{\JordanBlockStructureProc}{$A$, $\chi_A(\lambda)$, $f(\lambda)$, 
    $\cE$}
      \State{$\cV\gets\mbox{\KrylovGSProc($A$, $\chi_A(\lambda)$, $f(\lambda)$, 
      $\cE$)}$}
      \Comment{\Cref{alg:KrylovGS}}
      \State{$\{\tilde{\cV},\lbar\}\gets\mbox{\ExtKrylovGSProc($A$, $f(\lambda)$, $\cV$)}$}
      \Comment{\Cref{alg:ExtKrylovGS-matrix}}
      \State{$\cC\gets\mbox{\JordanBlockStructureMainProc($f(A)$, 
      $\tilde{\cV}=\bigcup_{\ell=1}^{\lbar}\cV^{(\ell)}$,
      $m_1$)}$}
      \Comment{\Cref{alg:jordan-blocks-structure-main-alt}}
      \State{\Return $\cC$}
    \EndFunction
  \end{algorithmic}
\end{algorithm}

\section{Complexity analysis}
\label{sec:complexity-analysis}

In this section, we analyze the time complexity of
\Cref{alg:jordan-blocks-structure} in terms of the number of arithmetic operations over $\K$.

Let $n$ be the size of the matrix $A$,
$m$ be the multiplicity of $f(\lambda)$ in $\chi_A(\lambda)$ (see \cref{eq:chara-pol}),
$d=\deg f(\lambda)$,
$r=|\cV|$ (see \Cref{alg:KrylovGS}),
and $\lbar$ be the maximum size of the Jordan blocks associated to the root of
$f(\lambda)$ (see \Cref{sec:jordan-krylov-basis}).
Furthermore, assume that $g(\lambda)$ is squarefree and has 
$q$ irreducible factors. 
The complexity of \Cref{alg:jordan-blocks-structure} is analyzed as follows.
\begin{itemize}
  \item Computation of \Cref{alg:KrylovGS} is essentially included in that of 
  minimal annihilating polynomials \cite[Remark 8]{taj-oha-ter2022}. 
  Furthermore, the complexity of computing the minimal annihilating polynomials
  is known to be $O(n^3q^2r)$ for the deterministic algorithm and
  $O(n^2q^2r)$ for the randomized algorithm \cite[Proposition 21]{taj-oha-ter2022}.
  \item The complexity of \Cref{alg:ExtKrylovGS-matrix} is $O(\lbar n^2r)$ plus the computing time of $f(A)$.
    This is because the while loop in
    \Cref{alg:ExtKrylovGS-matrix} is executed at most $\lbar$ times,
    and in each loop, the multiplication of $f(A)$ and
    the column reduction of the matrix $[\cV']$ is executed
    in $O(n^2r)$ time.
  \item The complexity of \Cref{alg:jordan-blocks-structure-main-alt}
    is $O(dn^2)$ plus the complexity of 
    \Cref{alg:jordan-blocks-structure-loop-alt}.
    This is because calculating matrices $S$ and $W$ is done in 
    $O(dn^2)$ time.
  \item The complexity of \Cref{alg:jordan-blocks-structure-loop-alt} is
    $O(\lbar)$ times the complexity of \Cref{alg:Jordan-Krylov-elim-ell-alt}, since, 
    in the worst case, the while loop in \Cref{alg:jordan-blocks-structure-loop-alt}
    is executed $\lbar$ times.
  \item The complexity of \Cref{alg:Jordan-Krylov-elim-ell-alt} is
    $O(nr+dn^2)$.
    This is because the simultaneous column reduction of the augmented matrices
    $[W\mid\v']$ and $[S_{\ell}\mid\v]$ can be done in $O(nr)$ time,
    and the computation of $\cL_{A,d}(\r)$ and $\cL_{A,d}(\r')$ can be done in $O(dn^2)$ time.
    Thus, the complexity of \Cref{alg:jordan-blocks-structure-loop-alt} is
    $O(\lbar nr+d\lbar n^2)=O(d\lbar n^2)$.
    Since we have $d\lbar\le r$, the total complexity is $O(n^2 r)$.
\end{itemize}
Thus, the total complexity of \Cref{alg:jordan-blocks-structure} is
$O(n^3q^2r)$ with the deterministic algorithm, or  
$O(n^2q^2r)$ with the randomized algorithm, for computing minimal annihilating polynomials.

\section{Experiments}
\label{sec:experiments}

In this section, we show the results of the experiments of 
\Cref{alg:jordan-blocks-structure}.
For the given matrix $A$, 
we have compared the computing time and memory usage
with \Cref{alg:jordan-blocks-structure}
with those of {\JKelim} down to rank $1$.
More precisely, we have compared the computing time of the following methods:
\begin{enumerate}
  \item Simple {\JKelim},
  \item Simple execution of \Cref{alg:jordan-blocks-structure},
  \item \Cref{alg:jordan-blocks-structure} with {\JKelim} in a matrix form 
  at the beginning of the {\JKelim} at each rank.
\end{enumerate}
Furthermore, in each method, we have compared the computing time 
and the memory usage with and 
without a \emph{preprocessing}, which consists of 
the column reduction of the {\KrylovGS} at the 
beginning of the {\JKelim} at each rank.
So, we have compared six methods in total.

The input matrix $A$ is given as
\[
  A=
  P^{-1}
  \begin{pmatrix}
    C(f_1) & & & O\\
    &  C(f_2) & \\
    & & \ddots & \\
    O & & & C(f_q)
  \end{pmatrix}
  P,
\]
where $C(f_i)$ is the companion matrix of $f_i(\lambda)$ with 
$f_i$ is a monic irreducible polynomial over $\Z$, 
and $P$ is a permutation matrix over $\Z$.

To evaluate the performance of our proposed method, we conducted analogous computations using the computer algebra system Maple and compared the respective computation times.
Within the Maple environment, the ``JordanForm'' and ``FrobeniusForm'' functions from the LinearAlgebra package were executed using their default settings.

The experiments were performed in the following environment:
Intel Xeon Silver 4210
\linebreak
at 2.20 GHz, RAM 256 GB, Linux 5.4.0 (SMP), Asir Version 20210326,
\linebreak
and Maple 2021 (\cite{maple2021}).
% Apple M1 Max up to 3.2 GHz, RAM 32 GB, macOS 12.2.1, 
% Risa/Asir version 20230315.

\subsection{A matrix with a Jordan blocks of size $1$ and $4$}
\label{sec:example-jordan-blocks-size-1-4}

In the first experiment, the matrix $A$ has
one Jordan block of size $4$ and eight Jordan blocks of size $1$,
both of which are associated to the same eigenvalue
and the characteristic polynomial of $A$ has one irreducible factor.
Thus, the structure of $A$ is represented as $\{8,0,0,1\}$.
The degree of the irreducible factor $f_1(\lambda)$ has changed as
$4,8,12,16,20$, so that the size of the matrix $A$ has changed as
$48,96,144,192,240$, respectively.

\Cref{tab:computing-time-jordan-blocks-size-1-4-simplejkelim} shows the computing time 
(in seconds) of the experiment with simple {\JKelim}.
Hereafter, the table is divided into two parts: the first part (the rows with the ``W/o preprocessing'' column)
shows the computing time for the method without preprocessing 
(without the column reduction of the {\KrylovGS} at the beginning of the {\JKelim} at each rank),
and the second part (the rows with the ``With preprocessing'' column) shows the computing time for the method with preprocessing. 
For each size of the matrix $A$, the computing time shows as follows: 
``$f_1(A)$'' shows the computing time of $f_1(A)$, 
``AnnihPol'' shows the computing time of the minimal annihilating polynomial
$g_i(\lambda)$ (see \Cref{alg:KrylovGS}),
``KrylovGS'' shows the computing time of the {\KrylovGS},
``Preprocessing'' shows the computing time of the preprocessing, 
or the column reduction of {\KrylovGS},
``JKElim'' shows the computing time of the {\JKelim}.
In computing time, $a$E$b$ means $a\times 10^b$ seconds.
Otherwise, the computing time is rounded to $0.01$ seconds.

Note that, in \Cref{tab:computing-time-jordan-blocks-size-1-4-simplejkelim},
with preprocessing, the computing time of {\JKelim} is reduced to less than half of the computing time without preprocessing.

\Cref{tab:computing-time-jordan-blocks-size-1-4-alg6} shows the computing time of the experiment with a simple execution of \Cref{alg:jordan-blocks-structure}.
Comparing with the computing time with simple {\JKelim} in
\Cref{tab:computing-time-jordan-blocks-size-1-4-simplejkelim}, 
while the computing time of $f_1(A)$ and {\JKelim} remains almost the same,
the computing time of the minimal annihilating polynomial is much reduced.
Furthermore, with preprocessing, the computing time of {\JKelim} is reduced to less than half of the computing time without preprocessing.

\Cref{tab:computing-time-jordan-blocks-size-1-4-alg6-matrix} shows the computing time of the experiment with the execution of \Cref{alg:jordan-blocks-structure} with {\JKelim} in a matrix form at the beginning of the {\JKelim} at each rank. 
Compared with the computing time in \Cref{tab:computing-time-jordan-blocks-size-1-4-alg6},
the computing time of the {\JKelim} is much reduced. This shows 
the effectiveness of the {\JKelim} in a matrix form.

\Cref{tab:computing-time-jordan-blocks-size-1-4-maple} shows the computing time of the experiment with Maple.
It seems that the computing time of the Jordan form and the Frobenius form in Maple is proportional to $n^3$, where $n$ is the size of the matrix.

\subsection{A matrix with a Jordan blocks of size $2$ and $10$}
\label{sec:example-jordan-blocks-size-2-10}

In the experiment in this subsection, the matrix $A$ has
one Jordan block of size $2$ and $10$, 
both of which are associated to the same eigenvalue, and the characteristic polynomial of $A$ has one irreducible factor.
Thus, the structure of $A$ is represented as $\{0,1,0,0,0,0,0,0,0,1\}$.
The degree of the irreducible factor $f_1(\lambda)$ has changed as
$4,8,12,16,20$, so that the size of the matrix $A$ has changed as
$48,96,144,192,240$, respectively.
\Cref{tab:computing-time-jordan-blocks-size-2-10-simplejkelim} shows the computing time 
(in seconds) of the experiment with simple {\JKelim}.

\Cref{tab:computing-time-jordan-blocks-size-2-10-alg6} shows the computing time of the experiment with a simple execution of \Cref{alg:jordan-blocks-structure}.
Compared with the computing time in \Cref{tab:computing-time-jordan-blocks-size-2-10-simplejkelim},
the computing time of the minimal annihilating polynomial is reduced. 
However, the computing time of the {\JKelim} remains almost the same; thus, the computing time of the preprocessing is not reduced so much.

\Cref{tab:computing-time-jordan-blocks-size-2-10-alg6-matrix} shows the computing time of the experiment with the execution of \Cref{alg:jordan-blocks-structure} with {\JKelim} in a matrix form at the beginning of the {\JKelim} at each rank. 
Compared with the computing time in \Cref{tab:computing-time-jordan-blocks-size-2-10-alg6},
the computing time of the {\JKelim} is much reduced. This shows 
the effectiveness of the {\JKelim} in a matrix form, as in the previous subsection.

\Cref{tab:computing-time-jordan-blocks-size-2-10-maple} shows the computing time of the experiment with Maple.
While the proposed algorithm with naive Jordan-Krylov elimination has longer computation times compared to Maple, the application of \Cref{alg:jordan-blocks-structure} and the Jordan-Krylov elimination in a matrix form improves computational efficiency.

\subsection{A matrix with a Jordan blocks of size $6$}
\label{sec:example-jordan-blocks-size-6}

In the experiment in this subsection, the matrix $A$ has
two Jordan blocks of size $6$, 
and the characteristic polynomial of $A$ has one irreducible factor.
Thus, the structure of $A$ is represented as $\{0,0,0,0,0,2\}$.
The degree of the irreducible factor $f_1(\lambda)$ has changed as
$4,8,12,16,20$, so that the size of the matrix $A$ has changed as
$48,96,144,192,240$, respectively.
\Cref{tab:computing-time-jordan-blocks-size-6-simplejkelim} shows the computing time 
(in seconds) of the experiment with simple {\JKelim}.
Note that the computing time of the minimal annihilating polynomial dominates
total computing time.

\Cref{tab:computing-time-jordan-blocks-size-6-alg6} shows the computing time of the experiment with a simple execution of \Cref{alg:jordan-blocks-structure}.
Compared with the computing time in \Cref{tab:computing-time-jordan-blocks-size-6-simplejkelim},
the computing time of the minimal annihilating polynomial is drastically reduced, which 
makes the total computing time much reduced.

\Cref{tab:computing-time-jordan-blocks-size-6-alg6-matrix} shows the computing time of the experiment with the execution of \Cref{alg:jordan-blocks-structure} with {\JKelim} in a matrix form at the beginning of the {\JKelim} at each rank. 
Compared with the computing time in \Cref{tab:computing-time-jordan-blocks-size-6-alg6},
the computing time of the {\JKelim} is increased inversely.

We also note that, in \Cref{tab:computing-time-jordan-blocks-size-6-simplejkelim,tab:computing-time-jordan-blocks-size-6-alg6,tab:computing-time-jordan-blocks-size-6-alg6-matrix},
the effect of preprocessing does not appear, and, on the contrary, the computing time of the {\JKelim} is increased.

\Cref{tab:computing-time-jordan-blocks-size-6-maple} shows the computing time of the experiment with Maple.
We see that, in the proposed algorithm, the Jordan-Krylov elimination is significantly efficient,
and that the introduction of \Cref{alg:jordan-blocks-structure} makes a big improvement in computational efficiency.

\subsection{A matrix with more than one irreducible factor in the characteristic polynomial}
\label{sec:example-multiple-irreducible-factors}

In the experiment in this subsection, the characteristic polynomial of $A$ has two irreducible factors: $f_1(\lambda)$ and $f_2(\lambda)$ of the equal degree.
The structure of $A$ associated to the root of $f_1(\lambda)$ is represented as $\{4,0,0,1\}$,
and the structure of $A$ associated to the root of $f_2(\lambda)$ is represented 
as $\{2\}$.
The degree of the irreducible factor $f_1(\lambda)$ has changed as
$4,8,12,16,20$, so that the size of the matrix $A$ has changed as
$48,96,144,192,240$, respectively.
\Cref{tab:computing-time-jordan-blocks-size-1-4-2-2-simplejkelim} shows the computing time 
(in seconds) of the experiment with simple {\JKelim}.
Note that the computing time of the minimal annihilating polynomial dominates
total computing time.

\Cref{tab:computing-time-jordan-blocks-size-1-4-2-2-alg6} shows the computing time of the experiment with a simple execution of \Cref{alg:jordan-blocks-structure}.
Compared with the computing time in 
\Cref{tab:computing-time-jordan-blocks-size-1-4-2-2-simplejkelim},
the computing time of the minimal annihilating polynomial is drastically reduced
as approximately from $1/3$ to $1/6$, which makes the total computing time much reduced.
On the other hand, the computing time of the {\JKelim} is not reduced very much so that
it dominates the total computing time.

\Cref{tab:computing-time-jordan-blocks-size-1-4-2-2-alg6-matrix} shows the computing time of the experiment with the execution of \Cref{alg:jordan-blocks-structure} with {\JKelim} in a matrix form at the beginning of the {\JKelim} at each rank.
Compared with the computing time in \Cref{tab:computing-time-jordan-blocks-size-1-4-2-2-alg6},
while the computing time of the minimal annihilating polynomial remains almost the same,
the computing time of the {\JKelim} is reduced as approximately from $1/4$ to $1/5$, 
which reduces the total computing time. 
This shows the effectiveness of the {\JKelim} in a matrix form.

We also note that, in \Cref{tab:computing-time-jordan-blocks-size-1-4-2-2-simplejkelim,tab:computing-time-jordan-blocks-size-1-4-2-2-alg6,tab:computing-time-jordan-blocks-size-1-4-2-2-alg6-matrix},
the preprocessing reduces the computing time in all the methods.

\Cref{tab:computing-time-jordan-blocks-size-1-4-2-2-maple} shows the computing time of the experiment with Maple.
We see that, in the proposed algorithm,
the introduction of \Cref{alg:jordan-blocks-structure} and the matrix form of the Jordan-Krylov elimination
makes the computation faster than Maple for the first time.
In this experiment, computing time for the case of $\text{Size}(A)=192$ was extremely large.
Although the exact reason is unknown, it is necessary to investigate the cause of this phenomenon in the future.

\subsection{A matrix with many irreducible factors in the characteristic polynomial}
\label{sec:example-many-irreducible-factors}

In the experiment in this subsection, the characteristic polynomial of $A$ has six irreducible factors: $f_1(\lambda),\dots,f_6(\lambda)$ of the equal degree.
The structure of $A$ associated to the root of $f_1(\lambda)$ is represented as 
$\{0,0,0,1,0,1\}$, and the structure of $A$ associated to the root of 
$f_2(\lambda),\dots,f_6(\lambda)$ is represented as $\{0,1\}$.
\Cref{tab:computing-time-jordan-blocks-size-4-6-2-2-2-2-2-simplejkelim} shows the computing time 
(in seconds) of the experiment with simple {\JKelim}.
Note that the computing time of {\JKelim} and the minimal annihilating polynomial dominates
total computing time.

\Cref{tab:computing-time-jordan-blocks-size-4-6-2-2-2-2-2-alg6} shows the computing time of the experiment with a simple execution of \Cref{alg:jordan-blocks-structure}, and
\Cref{tab:computing-time-jordan-blocks-size-4-6-2-2-2-2-2-alg6-matrix} shows the computing time of the experiment with the execution of \Cref{alg:jordan-blocks-structure} with {\JKelim} in a matrix form at the beginning of the {\JKelim} at each rank.
Compared to the computing time with a simple execution of the {\JKelim}, we have three observations. First, the computing time of the minimal annihilating polynomial
is hardly decreased, 
% from that with a simple execution of the {\JKelim},
which is similar between the two algorithms and which shows little difference with or without preprocessing.
Second, the computing time of the {\JKelim} is much reduced, which shows little difference with or without preprocessing.
Third, the computing time of the {\KrylovGS} is increased, which shows little difference with or without preprocessing.
In summary, computing time is not reduced so much with \Cref{alg:jordan-blocks-structure} or its modification.

\Cref{tab:computing-time-jordan-blocks-size-4-6-2-2-2-2-2-maple} shows the computing time of the experiment with Maple.
This example demonstrates that the proposed algorithm is particularly effective when the characteristic polynomial contains a large number of irreducible factors.

\section{Concluding remarks}
\label{sec:concluding-remarks}

In this paper, we have proposed an exact and efficient algorithm for computing the structure of 
a linear transformation. 
By focusing on the structure rather than the full Jordan chains, 
we successfully reduced computation in previous methods. 
% The effectiveness of our proposed algorithm has been demonstrated through experiments.

Through experiments, we demonstrated that the proposed method outperforms our original approaches
used for computing generalized eigenspaces as well as Maple’s built-in functions, especially when the characteristic polynomial contains multiple irreducible factors.

% In future work, it is desirable to explore potential applications for the proposed method and to develop efficient computational methods to those applications.

%\bibliographystyle{elsarticle-num-names-alpha}
% \bibliographystyle{elsarticle-harv}
%\bibliographystyle{plain}
\bibliographystyle{ACM-Reference-Format}

\bibliography{terui-e}

\begin{table}
  % \centering
  \caption{Computing time (in seconds) of the experiments in 
  \Cref{sec:example-jordan-blocks-size-1-4}.}
  \subcaption{With simple {\JKelim}.}
  \label{tab:computing-time-jordan-blocks-size-1-4-simplejkelim}
  % data: 20231203 / 411111111-02 fermat 01-1, 2-07-1, 2-07-2
  \begin{tabular}{l|r|rrrrr|r}
    \hline
    Preprocessing & Size($A$) & $f_1(A)$ & AnnihPol & KrylovGS & Preprocessing & JKElim & Total\\
    \hline
    W/o preprocessing & 48 & 0.08	& 0.38	& 2.17E$-4$	& --- & 0.68 & 1.13 \\
    & 96 & 1.18	& 5.22	& 4.29E$-4$	& --- & 4.18 & 10.58\\
    & 144 & 6.05	& 24.39	& 6.75E$-4$	& --- & 17.96 & 48.40\\
    & 192 & 25.07	& 85.32	& 9.69E$-4$	& --- & 58.75 & 169.14\\
    & 240 & 58.59	& 229.73	& 1.26E$-3$	& --- & 173.27 & 461.58\\
    \hline
    With preprocessing & 48 & 0.05 & 0.26 & 2.19E$-4$ & 0.13 & 0.37 & 0.81\\
    & 96 & 1.01 & 4.09 & 4.21E$-4$ & 0.57 & 2.52 & 8.19\\
    & 144 & 5.32 & 19.81 & 6.49E$-4$ & 1.45 & 9.18 & 35.76\\
    & 192 & 17.92	& 66.00 & 8.97E$-4$ & 3.11 & 24.23 & 111.26\\
    & 240 & 47.91 & 168.62 & 1.21E$-3$ & 5.63 & 67.99 & 290.15\\
    \hline
  \end{tabular}
% \end{table}
  \smallskip

% \begin{table}
%   \centering
  % \caption{Computing time (in seconds) of the experiment in
  % \Cref{sec:example-jordan-blocks-size-1-4}
  % with simple execution of \Cref{alg:jordan-blocks-structure}.}
  % data: 20231203 / 411111111-02 fermat 01-1, 2-07-1, 2-07-2
  \subcaption{With simple execution of \Cref{alg:jordan-blocks-structure}.}
  \label{tab:computing-time-jordan-blocks-size-1-4-alg6}
  \begin{tabular}{l|r|rrrrr|r}
    \hline
    Preprocessing & Size($A$) & $f_1(A)$ & AnnihPol & KrylovGS & Preprocessing & JKElim & Total\\
    \hline
    W/o preprocessing 
    & 48 & 0.07 & 0.06 & 2.0E$-4$ & --- & 0.34 & 0.47\\
    & 96 & 1.22 & 0.53 & 3.0E$-4$ & --- & 3.22 & 4.97\\
    & 144 & 6.30 & 1.61 & 5.0E$-4$ & --- & 17.55 & 25.47\\
    & 192 & 29.75 & 4.85 & 7.0E$-4$ & --- & 51.81 & 86.42\\
    & 240 & 72.63 & 10.42 & 8.0E$-4$ & --- & 169.43 & 252.48\\
    \hline
    With preprocessing 
    & 48 & 0.05 & 0.05 & 2.0E$-4$ & 0.11 & 0.22 & 0.43\\
    & 96 & 1.04 & 0.49 & 3.0E$-4$ & 0.51 & 1.62 & 3.67\\
    & 144 & 5.67 & 1.61 & 5.0E$-4$ & 1.34 & 6.67 & 15.29\\
    & 192 & 20.61 & 4.48 & 7.0E$-4$ & 2.92 & 19.81 & 47.82\\
    & 240 & 67.29 & 9.83 & 8.0E$-4$ & 5.47 & 67.61 & 150.20\\
    \hline
  \end{tabular}
% \end{table}
  \smallskip

% \begin{table}
%   \centering
  \subcaption{With 
  % \Cref{alg:jordan-blocks-structure} and
   {\JKelim} in a matrix form.}
  %  at the beginning of the {\JKelim} at each rank.}
  \label{tab:computing-time-jordan-blocks-size-1-4-alg6-matrix}
  \begin{tabular}{l|r|rrrrr|r}
    \hline
    Preprocessing & Size($A$) & $f_1(A)$ & AnnihPol & KrylovGS & Preprocessing & JKElim & Total\\
    \hline
    W/o preprocessing 
    & 48 & 0.07 & 0.06 & 1.22E$-3$ & --- & 0.25 & 0.38\\
    & 96 & 1.12 & 0.51 & 3.0E$-4$ & --- & 1.73 & 3.36\\
    & 144 & 6.11 & 1.60 & 5.0E$-4$ & --- & 7.72 & 15.43\\
    & 192 & 27.59 & 5.28 & 7.0E$-4$ & --- & 19.38 & 52.25\\
    & 240 & 71.18 & 11.53 & 8.0E$-4$ & --- & 51.54 & 134.26\\
    \hline
    With preprocessing 
    & 48 & 0.06 & 0.05 & 1.98E$-4$ & 0.12 & 0.19 & 0.42\\
    & 96 & 1.06 & 0.49 & 3.0E$-4$ & 0.52 & 1.26 & 3.33\\
    & 144 & 5.65 & 1.65 & 5.0E$-4$ & 1.36 & 4.20 & 12.85\\
    & 192 & 24.11 & 5.37 & 7.0E$-4$ & 3.09 & 11.36 & 43.92\\
    & 240 & 65.30 & 11.15 & 8.0E$-4$ & 5.59 & 25.11 & 107.15\\
    \hline
  \end{tabular}
  \smallskip

  \subcaption{Experiments with Maple.}
  \label{tab:computing-time-jordan-blocks-size-1-4-maple}
  % data: experiment/2023-11/411111111/411111111-02/maple/maple-4111111111-02-fermat.log
  \begin{tabular}{r|r|r}
    \hline
    Size($A$) & JordanForm & FrobeniusForm \\
    \hline
    48 & 0.38 & 0.31 \\
    96 & 3.93 & 3.57 \\
    144 & 17.79 & 16.94 \\
    192 & 48.87 & 47.62 \\
    240 & 165.00 & 162.00 \\
    \hline
  \end{tabular}
\end{table}

\begin{table}
  \caption{Computing time (in seconds) of the experiments in 
  \Cref{sec:example-jordan-blocks-size-2-10}.}
  \subcaption{With simple {\JKelim}.}
  \label{tab:computing-time-jordan-blocks-size-2-10-simplejkelim}
  % data: 20231203 / 102-02 fermat 01-1, 2-07-1, 2-07-2
  \begin{tabular}{l|r|rrrrr|r}
    \hline
    Preprocessing & Size($A$) & $f_1(A)$ & AnnihPol & KrylovGS & Preprocessing & JKElim & Total\\
    \hline
    W/o preprocessing 
    & 48 & 0.07 & 1.00 & 2.09E$-4$ & --- & 1.58 & 2.65\\
    & 96 & 1.08 & 13.58 & 3.99E$-4$ & --- & 17.92 & 32.59\\
    & 144 & 5.29 & 94.20 & 6.32E$-4$ & --- & 126.45 & 225.93\\
    & 192 & 22.52 & 264.83 & 9.41E$-4$ & --- & 445.43 & 732.77\\
    & 240 & 48.47 & 763.96 & 1.22E$-3$ & --- & 1986.88 & 2799.31\\
    \hline
    With preprocessing
    & 48 & 0.05 & 0.72 & 2.14E$-4$ & 0.23 & 0.71 & 1.71\\
    & 96 & 1.03 & 10.82 & 4.35E$-4$ & 2.59 & 9.97 & 24.40\\
    & 144 & 5.33 & 56.08 & 6.60E$-4$ & 5.32 & 41.26 & 107.98\\
    & 192 & 17.43 & 171.54 & 9.10E$-4$ & 22.32 & 177.03 & 388.33\\
    & 240 & 43.84 & 490.52 & 1.26E$-3$ & 31.19 & 557.77 & 1123.33\\
    \hline
  \end{tabular}
  \smallskip

  \subcaption{With simple execution of \Cref{alg:jordan-blocks-structure}.}
  \label{tab:computing-time-jordan-blocks-size-2-10-alg6}
  % data: 20231203 / 102-02 fermat 01-1, 2-07-1, 2-07-2
  \begin{tabular}{l|r|rrrrr|r}
    \hline
    Preprocessing & Size($A$) & $f_1(A)$ & AnnihPol & KrylovGS & Preprocessing & JKElim & Total\\
    \hline
    W/o preprocessing 
    & 48 & 0.06 & 0.19 & 1.66E$-4$ & --- & 1.20 & 1.45\\
    & 96 & 1.08 & 1.46 & 1.27E$-4$ & --- & 16.17 & 18.71\\
    & 144 & 5.54 & 5.44 & 4.59E$-4$ & --- & 136.22 & 147.19\\
    & 192 & 20.30 & 14.20 & 6.53E$-4$ & --- & 456.58 & 491.08\\
    & 240 &	55.85 & 38.34 & 8.31E$-4$ & ---	& 2030.34 & 2124.53\\
    \hline
    With preprocessing
    & 48 & 0.05 & 0.16 & 1.64E$-4$ & 0.23 & 0.62 & 1.06\\
    & 96 & 0.98 & 1.48 & 3.24E$-4$ & 2.58 & 8.94 & 13.97\\
    & 144 & 5.28 & 5.20 & 4.80E$-4$ & 5.45 & 37.96 & 53.89\\
    & 192	& 17.22	& 12.18	& 6.50E$-4$	& 21.62	& 165.20 & 216.22\\
    & 240 & 44.44 & 33.61 & 8.16E$-4$ & 30.56 & 554.50 & 663.12\\
    \hline
  \end{tabular}
  \smallskip

  \subcaption{With 
  % \Cref{alg:jordan-blocks-structure} and 
  {\JKelim} in a matrix form.}
  %  at the beginning of the {\JKelim} at each rank.}
  \label{tab:computing-time-jordan-blocks-size-2-10-alg6-matrix}
  % data: 20231203 / 102-02 fermat 01-1, 2-07-1, 2-07-2
  \begin{tabular}{l|r|rrrrr|r}
    \hline
    Preprocessing & Size($A$) & $f_1(A)$ & AnnihPol & KrylovGS & Preprocessing & JKElim & Total\\
    \hline
    W/o preprocessing 
    & 48 & 0.06 & 0.17 & 1.81E$-4$ & --- & 1.00 & 1.23\\
    & 96 & 1.06 & 1.43 & 3.18E$-4$ & --- & 13.11 & 15.61\\
    & 144 & 5.45 & 5.49 & 4.69E$-4$ & --- & 64.35 & 75.29\\
    & 192 & 19.14 & 14.91 & 6.57E$-4$ & --- & 186.04 & 220.09\\
    & 240 & 68.44 & 50.01 & 8.77E$-4$ & --- & 596.33 & 714.78\\
    \hline
    With preprocessing
    & 48 & 0.05 & 0.16 & 2.10E$-4$ & 0.17 & 0.55 & 0.93\\
    & 96 & 1.07 & 1.45 & 3.54E$-4$ & 1.53 & 5.28 & 9.33\\
    & 144 & 5.90 & 6.99 & 5.22E$-4$ & 6.86 & 27.19 & 46.95\\
    & 192 & 26.33 & 22.30 & 7.14E$-4$ & 22.60 & 81.77 & 153.01\\
    & 240 & 65.52 & 52.24 & 8.89E$-4$ & 45.64 & 184.72 & 348.12\\
    \hline
  \end{tabular}
  \smallskip

  \subcaption{Experiments with Maple.}
  \label{tab:computing-time-jordan-blocks-size-2-10-maple}
  % data: experiment/2023-11/102/102-02/maple/maple-102-02-fermat.log
  \begin{tabular}{r|r|r}
    \hline
    Size($A$) & JordanForm & FrobeniusForm \\
    \hline
    48 & 0.88 & 0.80 \\
    96 & 17.37 & 17.03 \\
    144 & 138.00 & 138.00 \\
    192 & 433.80 & 429.60 \\
    240 & 1450.80 & 1452.00 \\
    \hline
  \end{tabular}
\end{table}

\begin{table}
  \caption{Computing time (in seconds) of the experiments in 
  \Cref{sec:example-jordan-blocks-size-6}.}
  \subcaption{With simple {\JKelim}.}
  \label{tab:computing-time-jordan-blocks-size-6-simplejkelim}
  % data: 20231203 / 66-02 fermat 01-1, 2-07-1, 2-07-2

  \begin{tabular}{l|r|rrrrr|r}
    \hline
    Preprocessing & Size($A$) & $f_1(A)$ & AnnihPol & KrylovGS & Preprocessing & JKElim & Total\\
    \hline
    W/o preprocessing 
    & 48 & 0.07 & 0.54 & 2.08E$-4$ & --- & 8.76E$-3$ & 0.55\\
    & 96 & 1.31 & 7.92 & 3.99E$-4$ & --- & 0.05 & 7.97\\
    & 144 & 6.19 & 40.96 & 6.41E$-4$ & --- & 0.15 & 41.11\\
    & 192 & 25.85 & 111.76 & 9.55E$-4$ & --- & 0.32 & 112.07\\
    & 240 & 70.10 & 409.74 & 1.29E$-3$ & --- & 1.10 & 410.84\\
    \hline
    With preprocessing
    & 48 & 0.05 & 0.42 & 2.34E$-4$	& 0.06 & 0.04 & 0.52\\
    & 96 & 1.06 & 6.55 & 4.24E$-4$ & 0.27 & 0.17 & 7.00\\
    & 144 & 5.84 & 33.96 & 6.57E$-4$ & 0.65 & 0.46 & 35.07\\
    & 192 & 19.70 & 102.28 & 9.21E$-4$ & 1.27 & 0.93 & 104.47\\
    & 240 & 55.91 & 374.35 & 1.21E$-4$ & 2.44 & 2.25 & 379.03\\
    \hline
  \end{tabular}
  \smallskip

  \subcaption{With simple execution of \Cref{alg:jordan-blocks-structure}.}
  \label{tab:computing-time-jordan-blocks-size-6-alg6}
  % data: 20231203 / 66-02 fermat 01-1, 2-07-1, 2-07-2
  \begin{tabular}{l|r|rrrrr|r}
    \hline
    Preprocessing & Size($A$) & $f_1(A)$ & AnnihPol & KrylovGS & Preprocessing & JKElim & Total\\
    \hline
    W/o preprocessing 
    & 48 & 0.06 & 0.12 & 1.82E$-4$ & --- & 5.30E$-3$ & 0.18\\
    & 96 & 1.24 & 0.91 & 3.09E$-4$ & --- & 0.04 & 2.19\\
    & 144	& 6.15	& 2.93	& 4.82E$-4$	& --- & 0.13 & 9.21\\
    & 192 & 22.53	& 7.69	& 6.38E$-4$	& --- & 0.29 & 30.52\\
    & 240 & 70.88 & 25.30 & 8.32E$-4$	& --- & 0.97 & 97.14\\
    \hline
    With preprocessing
    & 48 & 0.06 & 0.10 & 2.03E$-4$ & 0.06 & 0.04 & 0.25\\
    & 96 & 1.09 & 0.85 & 3.27E$-4$ & 0.28 & 0.17 & 2.38\\
    & 144 & 6.22 & 3.24 & 4.72E$-4$ & 0.70 & 0.44 & 10.60\\
    & 192 & 26.73 & 9.29 & 6.52E$-4$ & 1.45 & 0.97 & 38.44\\
    & 240 & 70.76 & 23.97 & 8.31E$-4$ & 2.47 & 2.15 & 99.36\\
    \hline
  \end{tabular}
  \smallskip

  \subcaption{With 
  % \Cref{alg:jordan-blocks-structure} and 
  {\JKelim} in a matrix form.}
  %  at the beginning of the {\JKelim} at each rank.}
  \label{tab:computing-time-jordan-blocks-size-6-alg6-matrix}
  % data: 20231203 / 66-02 fermat 01-1, 2-07-1, 2-07-2
  \begin{tabular}{l|r|rrrrr|r}
    \hline
    Preprocessing & Size($A$) & $f_1(A)$ & AnnihPol & KrylovGS & Preprocessing & JKElim & Total\\
    \hline
    W/o preprocessing 
    & 48 & 0.06 & 0.11 & 1.89E$-4$ & --- & 0.12 & 0.30\\
    & 96 & 1.22 & 0.88 & 3.12E$-4$ & --- & 0.92 & 3.02\\
    & 144 & 6.03 & 3.06 & 4.68E$-4$ & --- & 4.43 & 13.52\\
    & 192 & 25.41 & 8.29 & 6.52E$-4$ & --- & 11.76 & 45.46\\
    & 240 & 68.92 & 20.59 & 8.22E$-4$ & --- & 32.27 & 121.78\\
    \hline
    With preprocessing
    & 48 & 0.05 & 0.10 & 2.07E$-4$ & 0.06 & 0.12 & 0.33\\
    & 96 & 1.07 & 0.82 & 3.44E$-4$ & 0.26 & 0.90 & 3.05\\
    & 144 & 5.83 & 3.28 & 4.91E$-4$ & 0.66 & 3.63 & 13.40\\
    & 192 & 25.35 & 9.42 & 6.73E$-4$ & 1.27 & 10.54 & 46.58\\
    & 240 & 71.49 & 20.86 & 8.27E$-4$ & 2.14 & 32.71 & 127.20\\
    \hline
  \end{tabular}
  \smallskip

  \subcaption{Experiments with Maple.}
  \label{tab:computing-time-jordan-blocks-size-6-maple}
  % data: experiment/2023-11/66/66-02/maple/maple-66-02-fermat.log
  \begin{tabular}{r|r|r}
    \hline
    Size($A$) & JordanForm & FrobeniusForm \\
    \hline
    48 & 0.51 & 0.48 \\
    96 & 7.98 & 8.45 \\
    144 & 50.99 & 53.12 \\
    192 & 156.00 & 157.20 \\
    240 & 510.00 & 492.00 \\
    \hline
  \end{tabular}
\end{table}

\begin{table}
  \caption{Computing time (in seconds) of the experiments in 
  \Cref{sec:example-multiple-irreducible-factors}.}
  \subcaption{With simple {\JKelim}.}
  \label{tab:computing-time-jordan-blocks-size-1-4-2-2-simplejkelim}
  % data: 20231209 / 41111-11-11-01 fermat 01-1, 2-07-1, 2-07-2
  \begin{tabular}{l|r|rrrrr|r}
    \hline
    Preprocessing & Size($A$) & $f_1(A)$ & AnnihPol & KrylovGS & Preprocessing & JKElim & Total\\
    \hline
    W/o preprocessing 
    & 48 & 0.05 & 0.27 & 0.03	& --- & 0.35 & 0.68\\
    & 96 & 0.86 & 4.04 & 0.46 & --- & 3.05 & 8.41\\
    & 144	& 4.47 & 18.53 & 2.11 & --- & 11.35 & 36.47\\
    & 192	& 13.33 & 59.89 & 5.94 & --- & 52.63 & 131.80\\
    & 240	& 33.82 & 161.10 & 13.34 & --- & 114.46 & 322.72\\ 
    \hline
    With preprocessing
    & 48 & 0.03 & 0.19 & 0.02 & 0.07 & 0.17 & 0.48\\
    & 96 & 0.77	& 3.32 & 0.41 & 0.37 & 1.24 & 6.12\\
    & 144	& 4.09 & 15.90 & 1.95 & 1.28 & 4.38 & 27.60\\
    & 192 & 12.69 & 48.76 & 5.84 & 2.79 & 12.04 & 82.12\\
    & 240 & 31.96 & 132.48 & 13.18 & 6.08 & 45.40 & 229.11\\
    \hline
  \end{tabular}
  \smallskip

  \subcaption{With simple execution of \Cref{alg:jordan-blocks-structure}.}
  \label{tab:computing-time-jordan-blocks-size-1-4-2-2-alg6}
  % data: 20231209 / 41111-11-11-01 fermat 01-1, 2-07-1, 2-07-2
  \begin{tabular}{l|r|rrrrr|r}
    \hline
    Preprocessing & Size($A$) & $f_1(A)$ & AnnihPol & KrylovGS & Preprocessing & JKElim & Total\\
    \hline
    W/o preprocessing 
    & 48 & 0.04 & 0.09 & 0.03 & --- & 0.09 & 0.26\\
    & 96 & 0.88 & 0.88 & 0.54 & --- & 1.11 & 3.42\\
    & 144 & 4.39 & 3.63 & 2.32 & --- & 6.07 & 16.42\\
    & 192 & 14.13 & 9.40 & 6.87 & --- & 19.19 & 49.59\\
    & 240 & 36.69 & 25.51 & 15.83 & --- & 95.28 & 173.31\\
    \hline
    With preprocessing
    & 48 & 0.03 & 0.09 & 0.03 & 0.05 & 0.11 & 0.30\\
    & 96 & 0.79 & 0.80 & 0.48 & 0.26 & 0.86 & 3.20\\
    & 144 & 4.21 & 3.19 & 2.24 & 0.74 & 3.75 & 14.13\\
    & 192 & 13.30 & 8.58 & 6.68 & 1.58 & 11.34 & 41.48\\
    & 240 & 33.80 & 23.14 & 15.31 & 2.99 & 42.23 & 117.47\\
    \hline
  \end{tabular}
  \smallskip

  \subcaption{With 
  % \Cref{alg:jordan-blocks-structure} and 
  {\JKelim} in a matrix form.}
  %  at the beginning of the {\JKelim} at each rank.}
  \label{tab:computing-time-jordan-blocks-size-1-4-2-2-alg6-matrix}
  % data: 20231209 / 41111-11-11-01 fermat 01-1, 2-07-1, 2-07-2
  \begin{tabular}{l|r|rrrrr|r}
    \hline
    Preprocessing & Size($A$) & $f_1(A)$ & AnnihPol & KrylovGS & Preprocessing & JKElim & Total\\
    \hline
    W/o preprocessing 
    & 48 & 0.04 & 0.09 & 0.04 & --- & 0.06 & 0.23\\
    & 96 & 0.91 & 0.90 & 0.56 & --- & 0.45 & 2.83\\
    & 144 & 4.67 & 3.68 & 2.40 & --- & 2.10 & 12.86\\
    & 192 & 14.06 & 9.33 & 6.82 & --- & 5.91 & 36.13\\
    & 240 & 38.45 & 25.76 & 15.90 & --- & 20.81 & 100.92\\
    \hline
    With preprocessing
    & 48 & 0.04 & 0.07 & 0.03 & 0.04 & 0.08 & 0.26\\
    & 96 & 0.80 & 0.80 & 0.49 & 0.26 & 0.62 & 2.96\\
    & 144 & 4.27 & 3.22 & 2.25 & 0.72 & 2.15 & 12.61\\
    & 192 & 13.26 & 8.56 & 6.63 & 1.01 & 4.72 & 34.18\\
    & 240 & 37.37 & 24.24 & 15.68 & 3.16 & 14.76 & 95.21\\
    \hline
  \end{tabular}
  \smallskip

  \subcaption{Experiments with Maple.}
  \label{tab:computing-time-jordan-blocks-size-1-4-2-2-maple}
  % data: experiment/2023-11/41111-11-11/41111-11-11-01/maple/maple-41111-11-11-01-fermat.log
  \begin{tabular}{r|r|r}
    \hline
    Size($A$) & JordanForm & FrobeniusForm \\
    \hline
    48 & 1.38 & 0.33 \\
    96 & 3.74 & 3.38 \\
    144 & 22.17 & 20.87 \\
    192 & 838.20 & 817.20 \\
    240 & 171.60 & 166.20 \\
    \hline
  \end{tabular}
\end{table}

\begin{table}
  \caption{Computing time (in seconds) of the experiments in 
  \Cref{sec:example-many-irreducible-factors}.}
  \subcaption{With simple {\JKelim}.}
  \label{tab:computing-time-jordan-blocks-size-4-6-2-2-2-2-2-simplejkelim}
  % data: 20231209 / 64-22222-03 fermat 01-1, 2-07-1, 2-07-2
  \begin{tabular}{l|r|rrrrr|r}
    \hline
    Preprocessing & Size($A$) & $f_1(A)$ & AnnihPol & KrylovGS & Preprocessing & JKElim & Total\\
    \hline
    W/o preprocessing 
    & 40 & 3.77E$-03$ & 0.06 & 0.01 & --- & 0.03 & 0.10\\
    & 80 & 0.05 & 0.39 & 0.03 & --- & 0.25 & 0.72\\
    & 120 & 0.17 & 1.24 & 0.07 & --- & 0.95 & 2.43\\
    & 160 & 0.41 & 2.78 & 0.13 & --- & 2.80 & 6.12\\
    & 200 & 0.86 & 5.64 & 0.20 & --- & 7.47 & 14.16\\
    \hline
    With preprocessing
    & 40 & 2.99E$-03$ & 0.04 & 0.01 & 0.02 & 0.03 & 0.10\\
    & 80 & 0.04 & 0.30 & 0.02 & 0.08 & 0.22 & 0.66\\
    & 120 & 0.15 & 1.09 & 0.06 & 0.23 & 0.71 & 2.24\\
    & 160 & 0.39 & 2.65 & 0.12 & 0.49 & 1.75 & 5.40\\
    & 200 & 0.87 & 5.57 & 0.19 & 0.90 & 3.98 & 11.51\\
    \hline
  \end{tabular}
  \smallskip

  \subcaption{With simple execution of \Cref{alg:jordan-blocks-structure}.}
  \label{tab:computing-time-jordan-blocks-size-4-6-2-2-2-2-2-alg6}
  % data: 20231209 / 64-22222-03 fermat 01-1, 2-07-1, 2-07-2
  \begin{tabular}{l|r|rrrrr|r}
    \hline
    Preprocessing & Size($A$) & $f_1(A)$ & AnnihPol & KrylovGS & Preprocessing & JKElim & Total\\
    \hline
    W/o preprocessing 
    & 40 & 4.25E$-03$ & 0.06 & 0.02	& --- & 0.01 & 0.09\\
    & 80 & 0.04 & 0.39 & 0.17 & --- & 0.05 & 0.65\\
    & 120	& 0.16 & 1.27 & 0.56 & --- & 0.17 & 2.15\\
    & 160 & 0.41 & 2.88 & 1.32 & --- & 0.47 & 5.07\\
    & 200 & 0.85 & 5.75 & 2.65 & --- & 0.80 & 10.04\\
    \hline
    With preprocessing
    & 40 & 3.11E$-03$ & 0.05 & 0.02 & 0.01 & 0.01 & 0.08\\
    & 80 & 0.04 & 0.33 & 0.14 & 0.02 & 0.04 & 0.57\\
    & 120 & 0.15 & 1.13 & 0.51 & 0.06 & 0.14 & 1.99\\
    & 160 & 0.39 & 2.65 & 1.24 & 0.11 & 0.33 & 4.72\\
    & 200 & 0.84 & 5.62 & 2.61 & 0.20 & 0.71 & 9.98\\
    \hline
  \end{tabular}
  \smallskip

  \subcaption{With 
  % \Cref{alg:jordan-blocks-structure} and 
  {\JKelim} in a matrix form.}
  %  at the beginning of the {\JKelim} at each rank.}
  \label{tab:computing-time-jordan-blocks-size-4-6-2-2-2-2-2-alg6-matrix}
  % data: 20231209 / 64-22222-03 fermat 01-1, 2-07-1, 2-07-2
  \begin{tabular}{l|r|rrrrr|r}
    \hline
    Preprocessing & Size($A$) & $f_1(A)$ & AnnihPol & KrylovGS & Preprocessing & JKElim & Total\\
    \hline
    W/o preprocessing 
    & 40 & 5.05E$-03$ & 0.06 & 0.02 & --- & 0.01 & 0.10\\
    & 80 & 0.04 & 0.39 & 0.17 & --- & 0.05 & 0.65\\
    & 120 & 0.15 & 1.26 & 0.56 & --- & 0.15 & 2.13\\
    & 160 & 0.40 & 2.90 & 1.33 & --- & 0.38 & 5.01\\
    & 200 & 0.85 & 5.75 & 2.64 & --- & 0.62 & 9.86\\
    \hline
    With preprocessing

    & 40 & 3.02E$-03$ & 0.05 & 0.02 & 0.01 & 0.01 & 0.09\\
    & 80 & 0.04 & 0.33 & 0.14 & 0.03 & 0.05 & 0.58\\
    & 120 & 0.15 & 1.13 & 0.51 & 0.06 & 0.14 & 1.99\\
    & 160 & 0.39 & 2.64 & 1.23 & 0.11 & 0.31 & 4.68\\
    & 200 & 0.82 & 5.50 & 2.60 & 0.19 & 0.62 & 9.74\\
    \hline
  \end{tabular}
  \smallskip

  \subcaption{Experiments with Maple.}
  \label{tab:computing-time-jordan-blocks-size-4-6-2-2-2-2-2-maple}
  % data: experiment/2023-09/64-22222/64-22222-03/maple/maple-64-22222-03-fermat.log
  \begin{tabular}{r|r|r}
    \hline
    Size($A$) & JordanForm & FrobeniusForm \\
    \hline
    40 & 0.25 & 0.20 \\
    80 & 4.15 & 2.82 \\
    120 & 19.81 & 19.25 \\
    160 & 56.72 & 55.52 \\
    200 & 147.60 & 146.40 \\
    \hline
  \end{tabular}
\end{table}

\end{document}